\documentclass[aps,preprintnumbers,nofootinbib,superscriptaddress]{revtex4}

\usepackage{graphicx}
\DeclareGraphicsExtensions{.pdf}

\usepackage{amsfonts,amssymb,amsmath}
\usepackage{amsthm}
\usepackage{color} 
\newcommand{\comment}[1]{}

\theoremstyle{plain}
\newtheorem{theorem}{Theorem}

\theoremstyle{definition}

\usepackage{graphicx}
\DeclareGraphicsExtensions{.pdf}

\usepackage{amsfonts,amssymb,amsmath}
\usepackage{amsthm}
\usepackage{color}

\usepackage{amsfonts,amssymb,amsmath}
\usepackage{amsthm}
\usepackage{color} 
\usepackage{amsmath}
\usepackage{amsthm}
\usepackage{amssymb}
\usepackage{amsfonts}
\usepackage{bbm}
\usepackage{epsfig}
\usepackage{times}
\usepackage{color}
\usepackage{hyperref}

\newtheorem{assumption}[theorem]{Assumption}

\begin{document} 
\begin{titlepage}

\title{The Problem of Confirmation in the Everett Interpretation }

\author{Emily \surname{Adlam}} \affiliation{Centre for Quantum
  Information and Foundations, DAMTP, Centre for Mathematical
  Sciences, University of Cambridge, Wilberforce Road, Cambridge, CB3
  0WA, U.K.}  \affiliation{The University of Oxford}

\maketitle

\tableofcontents

\end{titlepage}

\Large
\textbf{Abstract} 
\vspace{4mm}

\normalsize

\vspace{6mm}

I argue that the Oxford school Everett interpretation is internally incoherent, because we cannot claim that in an Everettian universe the kinds of reasoning we have used to arrive at our beliefs about quantum mechanics would lead us to form true beliefs. I show that in an Everettian context, the experimental evidence that we have available could not provide empirical confirmation for quantum mechanics, and moreover that we would not even be able to establish reference to the theoretical entities of quantum mechanics. I then consider a range of existing Everettian approaches to the probability problem and show that they do not succeed in overcoming this incoherence.


\newpage

\begin{Large} 
\textbf{Introduction} 
\end{Large}

\vspace{4mm}

Proponents of the Oxford school Everett interpretation like to claim that their theory is what follows if we  `interpret the bare quantum formalism in a straightforwardly realist way.' \cite{Wallace2011}. But from the perspective of scientific realism, the question of whether we should accept the Everett interpretation is a question about what is true of the world, and it follows that we are justified in accepting the Everett interpretation only if we can claim that in an Everettian universe, the kinds of reasoning we have used to arrive at our beliefs about quantum mechanics would lead us to form true beliefs. In this article, I argue that the Oxford school Everett interpretation is not in fact capable of supporting such a claim. 

In section one, I give a brief introduction to the Everett interpretation. In sections two and three, I provide two independent arguments showing that reasoning based on experimental evidence could not be expected to lead to true beliefs in an Everettian universe: first, I argue that there is no experimental outcome which could serve to confirm Everettian quantum mechanics over any other theory according to which that outcome is possible, and then I demonstrate that according to a certain plausible view of reference,  reasoning from experimental evidence would not even enable us to establish reference to the entities invoked by the Everettian theory. 

There are a number of existing strategies that have been invoked to deal with concerns about probability and confirmation in the Everett interpretation. In the final section of this article, I consider three particularly popular approaches and demonstrate that they are not able to overcome the epistemic problems I have identified. I first discuss the branch counting approach and explain why it has been rejected even by the Oxford school Everettians themselves; I then turn to the argument that the Everett interpretation is no worse off than other probabilistic theories, and demonstrate that in fact, ordinary probabilistic theories have a built-in connection to truth which is lacking in the Everettian case. I also consider the decision theoretic approach pioneered by the Oxford school Everettians, arguing that in the Everettian context, the usual conceptual connection between belief which is rational in the decision-theoretic sense and truth fails to hold. 

Finally, I conclude that in order to defend the Everett approach from the points raised in this article, it would be necessary to adopt a set of philosophical theses which are in tension if not outright contradiction with one another, and that therefore it is not rational to believe that Everettian quantum mechanics is true.

\newpage

\section{The Everett Interpretation}  

One of the most striking features of quantum theory is that systems can be in superpositions of states normally thought to be mutually exclusive. Moreoever, the time-dependent Schrodinger's equation implies that quantum systems undergo linear, deterministic evolution. This means that in any measurement of a quantum system, if we treat the measuring instrument quantum mechanically and apply the unitary equations, we find that the instrument itself is predicted to go into a superposition of macroscopically distinct states. This is problematic, since measuring instruments always appear to us to be in a single determinate state. Worse, because we must always interact with our instruments in order to discover the results of measurements, it would seem that we ourselves should end up in superposition states, which is difficult to reconcile with our experience of being in a single determinate state at all times. Thus we have what is sometimes referred to as the `measurement problem' - the challenge of explaining how the quantum formalism can possibly represent the universe as we experience it \cite{Krips}. 

Many proposed interpretations of quantum mechanics respond to this problem by altering the formalism - for example, adding a `dynamical collapse' so that superpositions are always reduced to a single branch before measurement outcomes are observed \cite{Ghirardi}.
However, in 1957 Everett noted that although quantum mechanics predicts that observers should end up in superposition states, the observer subsystem always has a unique determinate state relative to each distinct state of the instrument subsystem, so if we regard each  observer-instrument composite as a single observer, we can understand why measurements always appear to have a unique outcome. Everett's `relative state' formalism has since undergone significant development, and several different variants have emerged, including the Many-Worlds \cite{DeWitt}, Many-Minds \cite{AlbertLoewer} and Many-Histories \cite{GellmannHartle} approaches. With the development of decoherence theory it has become possible to show directly from the standard formalism of quantum mechanics that macroscopically distinct branches of the wavefunction are almost entirely free from interference and evolve approximately classically (see Wallace \cite{Wallace2010Decoherence}), and this has given rise to what we may call the Oxford school approach (championed by Deutsch \cite{Deutsch1999}, Saunders \cite{SaundersWallace}, Wallace \cite{Wallace2012}, Greaves \cite{Greaves2004} and others), which claims that the `worlds' may be understood as emergent features of the structure produced by decoherence, ' rather than being distinct elements of the ontology. This view has the advantage that it does not ask us to alter or add to the existing mathematical formalism of quantum mechanics, so considerations of ontological economy would suggest that if it is empirically adequate, it should be preferred to any interpretation which does require the addition of further structure \cite{Wallace2011}. 

But the Oxford school Everett approach is not without difficulties - in particular, it is not entirely clear that we can legitimately regard the evidence we have for quantum mechanics as confirmation for an Everettian version of quantum mechanics. Notice that this difficulty does not concern the confirmation of the Everett interpretation over other proposed interpretations, since the Everett interpretation makes use of the standard predictive algorithm of quantum mechanics, and is therefore empirically equivalent to other interpretations which use the same algorithm, or at least, equivalent with regard to all the evidence we currently have (see Tegmark, 1998 \cite{Tegmark}). Rather, the difficulty is about the confirmation of the standard predictive algorithm itself: as Barrett argues, \cite{Barrett} `whatever formulation of quantum mechanics we end up with it ought to be one where we are able to tell a coherent story in the context of the theory that explains how we came to have reliable empirical evidence for its acceptance.' Therefore if the Oxford school Everett interpretation is to be at all plausible, we must be able to claim that the reasoning  which we have used to arrive at our beliefs about quantum mechanics would still be expected to lead to true beliefs even in an Everettian universe. I now present two arguments demonstrating that such a claim cannot in fact be made.

\newpage

\section{Everettian Confirmation}

One useful way of thinking about the status of evidence in an Everettian universe is in terms of self-locating facts, which are facts that `locate the agent at a certain time or location (or possibly with respect to other variables).' \cite{Bradley} Propositions about these sorts of facts have to be treated differently from other propositions; in particular, ordinary propositions are frequently analysed as functions from possible worlds into truth values, but this approach is untenable for propositions about self-locating facts, since for example, the proposition that it is twelve o'clock now does not have a unique truth value for each possible world, as we will still be occupying the same possible world an hour later when it is one o'clock. Thus Lewis \cite{Lewis} proposes accommodating self-locating facts by introducing `centered worlds,' which are possible worlds paired with individuals and times, so that propositions about self-locating facts can be analysed as functions from centered worlds into truth values.

It seems plausible that observing a measurement result in an Everettian universe should be viewed as a case of locating oneself within a possible world with respect to which branch one occupies - that is, it involves learning only a self-locating fact. This suggests that we should expand the definition of centered worlds so that they are now understood as possible worlds paired with individuals, times and also branches; thus in an Everettian universe, observed measurement results enable us to distinguish between centered worlds but not possible worlds. In particular, suppose that we carry out an experiment and observe a given outcome, and suppose for simplicity that we are considering only a set of theories which all entail that this result is a possible one. The following argument can be made:

\begin{enumerate}
\item If Everettian quantum mechanics is true, I learn only a self-locating fact when I observe the result of this experiment. 
\item \textbf{ The Relevance-Limiting Thesis:} It is never epistemically rational for an agent who learns only self-locating information to respond by altering a non-self-locating credence \cite{Titelbaum}. 
\end{enumerate}
\textbf{Subconclusion:} Either Everettian quantum mechanics is not true, or it is not epistemically rational for me to alter my non-self-locating credences in response to this experiment. 
\vspace{1mm}

\textbf{Conclusion:} There is no outcome of this experiment to which the epistemically rational response is to increase my credence in the proposition that Everettian quantum mechanics is true. 
\vspace{2mm} 

The reason the conclusion follows from the subconclusion is that the question of whether Everettian quantum mechanics is true is equivalent to the question of whether or not I am in a possible world which is governed by Everettian quantum mechanics, and thus it is a matter of non-self-locating belief. Moreoever, since this argument can be made about more or less any quantum measurement we like, it looks as if there is no experimental outcome which could serve to confirm Everettian quantum mechanics over any other theory according to which that outcome is possible. In what follows I will strengthen this conclusion by motivating and defending each of the argument's premises.

\subsection{Premise 1}

In order to motivate Premise 1, it is helpful to consider a comparable case in a non-branching universe, where a person is told that he is to be duplicated and the resulting two people to be placed in different locations. After the operation, when one of those duplicates determines his location, does he learn anything that is not merely a self-locating fact? One way to answer this question in the affirmative would be to postulate some kind of Cartesian Ego which ends up inhabiting one post-duplication organism rather than the other. The possible world in which the Ego enters organism A would then be distinct from the possible world in which the Ego enters organism B, and therefore when the post-duplication person associated with the Ego observed his location he would gain non-self-locating information enabling him to distinguish between these two possible worlds. This evidence would also be capable of providing confirmation for certain kinds of theories: for example, if we had alternative theories about the behaviour of Cartesian Egos which assigned different values to the probability of the Ego entering the different post-duplication organisms, then after duplication, if the person associated with the Ego observed location A rather than B, Bayesian conditionalization would lead him to increase his credence in the theory which attaches a higher probability to the Ego entering organism A \footnote{See Howson and Urbach, 1993, \cite{HowsonUrbach} for an account of Bayesian conditionalization.}.

But there are several things wrong with this story. First of all, the whole notion of a Cartesian Ego is highly suspect, and virtually untenable if we accept a functionalist account of consciousness. Moreoever, the story works only if the post-duplication organism not inhabited by the Cartesian Ego is not conscious at all: we cannot accept that it is conscious in the absence of an Ego, or that a second Ego is created for it, because then observing a location would not enable either post-duplication person to distinguish between the worlds where the Ego enters organism 1 or organism 2, since there would be no way for either of these two conscious beings to know whether they were associated with the (former) Ego. Thus unless we are willing to adopt a very radical view of the nature of consciousness, we must accept that the post-duplication people learn only a self-locating fact when they observe their location. 

The case of the branching Everettian agent is analogous. Assume first that the subject knows that Everettian quantum mechanics is true, and knows exactly which outcomes will occur when the measurement is carried out. Then observing a particular outcome involves learning only a self-locating fact, unless we postulate a Cartesian Ego which enters one branch rather than another when splitting occurs. But this requires us to add non-physical facts about this Cartesian Ego to the universe, and as Ismael points out, this strategy `isn't very satisfying as physics.'  \cite{Ismael}. Furthermore, the mere existence of such an Ego does not suffice; as in the non-branching case, we must also claim that organisms in branches not inhabited by an Ego are not conscious at all, otherwise the person observing the result would be incapable of ascertaining whether or not he was the person associated with the Ego. Thus we end up with a proliferation of mindless automata \footnote{For further exploration of this problem, see Albert and Loewer, 1988 \cite{AlbertLoewer}.} which clearly runs contrary to the functionalist approach to consciousness promoted by Oxford school Everettians \footnote{The Oxford school version of the Everett interpretation is based on a broadly physicalist picture and treats consciousness, along with other macroscopic properties, as emergent from the underlying ontology, in accordance with a functionalist view of mentality. See Wallace, 2003 \cite{Wallace2003} }. But if we reject the postulation of a Cartesian Ego, it seems that an Everettian agent, like a post-duplication agent, can learn only a self-locating fact when he observes a measurement outcome. \footnote{Indeed, the notion that Everettian measurement outcomes give us only self-locating information is accepted by many Oxford school Everettians. Wilson writes that `objective probability ... is  irreducibly self-locating,' and even suggets that this `dissolves ... worries about objective uncertainty and expectation.' \cite{Wilson} However if the Relevance-Limiting Thesis is accepted, the argument of section two demonstrates that no view on which probabilities are merely self-locating facts can be an adequate solution to the epistemic side of the probability problem. }

Of course the real situation is more complex than this, because we ourselves are not sure which theory of the world is correct, so the world we are actually in may or may not contain branching. And if we are in a world which is not branching, then experimental results do give us non-self-locating information, since the different possible outcomes correspond to different possible non-branching worlds and therefore observing an outcome tells us something about which world we are in, not merely where we are within a given world. This means that in real situations, once any branching theory is admitted as a possibility, we simply don't know whether we should think of experimental outcomes as giving us information about non-self-locating facts or not. Whether a particular measurement can distinguish between variants of the theory thus depends crucially on whether we are in a branching world or not, and this places us in the problematic epistemic predicament I have outlined. 

\subsection{Premise 2}

The Relevance-Limiting Thesis seems well-motivated from a physicalist point of view: if you are told in advance all the physical facts about a set of physical systems (e.g. brains) which are to be instantiated in several different places at a certain time, it seems obvious that you can learn nothing new when at that time you find yourself associated with one of those physical systems rather than another, since you know that all the other physical systems are having comparable experiences. However, it would be best if these intuitions could be supported by a general account of how we ought to update our beliefs where self-locating propositions are concerned. 

Our usual approach to belief updating is based on what Greaves refers to as minimal conditionalization, which `simply amounts to setting one's credence to zero on the proposition whose falsity one has just learnt, and renormalizing.' \cite{Greaves2007} For example, if we perform an experiment at time $t_1$ and observe outcome $E_1$, we learn that the proposition `$E_1$ does not occur at time $t_1$,' is false, and we should therefore eliminate all possible worlds in which $E_1$ does not occur at $t_1$. A straightforward extension of this strategy to the case of self-locating belief in a branching universe would appear to uphold the Relevance-Limiting Thesis: an observation which serves only to pinpoint our location in a given world does not enable us to eliminate any possible worlds, and therefore such an observation cannot change the values of our credences in any possible worlds. Indeed, the Relevance-Limiting Thesis is implied by any account of epistemically rational belief updating on which we are supposed to first assign credences to possible worlds and then somehow distribute those credences over the centered worlds corresponding to the possible worlds, such as the system proposed by Halpern and Tuttle \cite{HalpernTuttle} or Meacham's compartmentalized conditionalization \cite{Meacham}. 

However, there are alternative accounts of epistemically rational belief updating which do not uphold the Relevance-Limiting Thesis. Lewis \cite{Lewis}, for example, claims that Bayesian conditionalization can be applied straightforwardly to self-locating beliefs by replacing possible worlds with centered worlds; on this approach, the credence assigned to a possible world is the sum of the credences assigned to all of the centered worlds associated with it, so our credences in possible worlds may be altered by a change in our credences for centered worlds, and therefore learning purely self-locating information can alter our non-self-locating beliefs. But Bayesian conditionalization allows us only to eliminate possible worlds, not to add them, and therefore Lewis' approach cannot deal with cases where the number of centered worlds under consideration can change. The proposal must therefore be supplemented with a strategy for extending Bayesian conditionalization to such cases: for example, Halpern \cite{Halpern} discusses an approach based on Elga's work \cite{Elga}, in which we first assign to each centered world the prior probability assigned to the corresponding possible world, then normalize over all centered worlds. Variations on this scheme are possible, of course, but typically any updating strategy that does not uphold the Relevance-Limiting Thesis will involve some process of renormalizing over the whole set of centered worlds.
This leads to difficulties for such strategies, because the renormalization process favours possible worlds which correspond to a larger number of centered worlds, yet it is not reasonable that I should be forced to have high credence in some possible world $W_1$ purely because it happens to contain a large number of duplicated brains in states identical to the state of my own brain. It might seem that we could avoid this difficulty by suitable choice of priors: if our prior credence for each centered world contained in $W_1$ is sufficiently low, then our total credence for $W_1$ can be made as small as we like, no matter which updating approach is used. However, Meacham \cite{Meacham} points out that difficulties will still arise when the number of centered worlds increases over time: if we use the Elga extension, our credence in certain sorts of propositions will increase over time, eventually converging to certainty no matter how low the initial credence assigned to them, so we will inevitably come to believe these propositions no matter how unlikely we originally thought them to be \footnote{For example, let $P$ be the proposition `At time $t_1$ my brain is duplicated, and this process is then repeated many times,' and let my initial credence in $P$ be $p(0)$, such that $ 0 \leq p(0) \leq 1 $. At time $t_1$ new centered worlds associated with the putative new brains start coming into being; if we use Elga's updating strategy, the ratio of the credence associated with each such centered world to the credence attached to $ \neg P$ must be a constant $C$ such that $\frac{1 - p(t)}{p(t)} = C $, and the total credence attached to $P$ at $t$ is given by $N(t)p(t)$, where $N(t)$ is the number of minds that have been created at $t$. Since $N(t)p(t) + Cp(t) = 1$, the total credence attached to $P$ at $t$ is $\frac{1}{1 + \frac{C}{N(t)}}$, which approaches 1 as $N(t)$ becomes large.}. Since this is not the way that belief-updating ought to work, we have good reason to reject Elga's extension. 

In the Everettian context, Elga's extension is not acceptable in any case, since it assigns each centered world equal probability rather than weighting by mod-squared amplitudes. In order to claim that Everettian quantum theory may inherit the existing confirmation for quantum theory, it is necessary for the Everettian to choose an extension in which the initial credence for some centered world is equal to the product of the credence attached to the corresponding possible world and the mod-squared amplitude assigned to the relevant centered world by the theory governing that possible world, and where, on observing an outcome, an Everettian agent eliminates all centered worlds in which that outcome does not occur and then redistribute credences over the remaining centered worlds so as to keep the ratios between them the same while ensuring that they sum to one. Greaves refers to this scheme as `Extended Conditionalization' \cite{Greaves2007}. Clearly Extended Conditionalization is not consistent with the Relevance-Limiting Thesis: for example, if there is some possible world $W$ whose governing theory assigns very high mod-squared amplitude to the centered world in which outcome $O_1$ is observed, and we fail to see outcome $O_1$, centered worlds corresponding to that outcome will be eliminated in all possible worlds, and our overall credence in $W$ will decrease because the change will favour those worlds whose governing theories assign lower mod-squared amplitude to $O_1$. However, Extended Conditionalization, just like the Elga approach, involves renormalizing over a set of centered worlds, and thus it too is vulnerable to the problem of ballooning credences in cases where the number of centered worlds increases. An analog of the paradox described by Meacham can be given in the quantum case, with duplication replaced by branching and the Elga extension replaced by Extended Conditionalization: if we consider the proposition `At time $t_0$ my brain undergoes a quantum branching process, and this process is then repeated many times,' then Extended Conditionalization implies that our credence in this proposition must increase over time, eventually approaching certainty no matter how low our initial credence was. Since this is not the way belief updating should work, it looks as though Extended Conditionalization cannot be accepted as a rational belief-updating strategy, so the Relevance-Limiting Thesis is upheld. 

\subsubsection{The Appeal to Philosophy of Language}

The most obvious way for the Everettian to counter this argument would be to claim that the number of centered worlds never increases: for example, we might take the view that centered worlds should be characterised not as branches at times but as entire personal histories, so that the history which goes to brain $B_1$ is one centered world, the history which goes to brain $B_2$ is another, and so on. This would imply that centered worlds exist eternally rather than being time-indexed, so the problem of ballooning credences would not arise and Extended Conditionalization would be a viable updating strategy after all. 

But such an analysis requires us to accept that the centered worlds associated with Everettian branches are distinct even before branching has occurred. Saunders and Wallace \cite{SaundersWallace} attempt to motivate this sort of claim by appeal to considerations drawn from the philosophy of language. They suggest that the most perspicacious way to think about the different individuals involved in a branching situation is as continuants rather than instantaneous temporal slices, and argue that `if persons are continuants, we do better to attribute thoughts and utterances at t to continuants at t,' and therefore `in the the presence of branching ... there are two or more thoughts or utterances expressed at t, one for each of the continuants that overlap at that time.' Thus when a pre-branching subject utters something like `I will observe outcome $O_1$,' the utterance is equivalent to `I am (already, before the measurement) the continuant that observes outcome $O_1$,' and should be thought of as being uttered by two separate continuants, one of whom is right and one of whom is wrong. This would imply that even before the measurement there are two distinct centered worlds available for consideration, one associated with each continuant. 

A first objection to this approach has been raised by Peter Lewis, who points out that on the reductivist approach to personhood, `it is not a deep truth that there are two persons (before branching), but merely a convenient way of tracking branching entities across time.' \cite{PeterLewis} This reductivist approach is implicit in the argument of Saunders and Wallace, since their claim that it is most perspicacious to think about the individuals involved as continuants presupposes that we are free to choose from a number of possible descriptions. Yet this reductivist approach also implies that our beliefs about the physical universe should be independent of how we choose to think about personhood, because the truth about physical reality cannot be determined by our way of describing it. Thus if the reasoning process by which we have arrived at our beliefs about quantum mechanics makes sense in an Everettian universe only provided we accept a certain semantic analysis of what it is to be a person, we can have no good reason to expect that beliefs formed in this way would be true in an Everettian universe. 

Moreoever, it is not clear that the two continuants to which Saunders and Wallace refer are capable of playing the necessary role in the process of confirming a theory. The fact that an observation has been made can confirm one theory over another only if the probability of its being made is higher conditional on the truth of that theory than on the other; but if two theories both imply that a particular observation will certainly be made by someone, its probability conditional on either theory is equal to one, so the fact that the observation has been made cannot confirm either theory over the other. Only if we have some means of individuating a particular observer which is independent of the outcome that he sees can we treat him as a random sample from the total reference class of observers, and thus define nontrivial probabilities, conditional on each of the theories in turn, for the observation being made by \emph{this} particular observer (see Leslie \cite{Leslie} for an illustrative example \footnote{ In Leslie's example, a group of one hundred women is to be divided into subsets of 95 and 5, and the name `Heads' randomly assigned to one of these sets by flipping a coin, with the other group to be named `Tails.' If we have no other information, we should assign probability 50\% to the hypothesis that the Heads group is the bigger group. But if we receive the information that `Liz is in the Heads group,' we should assign 95\% probability to the hypothesis that the Heads group is the bigger group, because we now have the information that some particular observer is in the Heads group. Moreover, our credences obviously should not change if we merely receive the information that `One of the women in the Heads group is in the Heads group,' - it is crucial that the means of denoting the observer does not depend on the outcome of the observation.}). Thus when we are comparing two different branching world hypotheses, both of which imply that a given measurement will have two outcomes which will be observed by two continuants on two separate branches, the fact that one of the outcomes is observed cannot confirm either theory over the other unless we have some means of singling out one of the continuants, thus enabling us to view that continuant as a random sample from our reference class of two continuants. But as Peter Lewis points out, `to say that the state has branches is just to say that it can be written as a sum of more-or-less independent terms, where each term is taken as a description of a state of affairs ... so the state of affairs \emph{is} the branch ... it makes no sense to conceive of the same state of affairs in a different branch.' In the same way, to say that there are two continuants present before branching is just to say that we can write two separate histories for observer subsystems, so the notion that there are two continuants present before branching depends on the claim that identity conditions for continuants are given by specifying their entire history, which means that a continuant necessarily sees the outcome that it does see. A continuant individuated in this way certainly cannot be viewed as a random sample, because the claim that such a continuant sees a certain outcome is equivalent to the claim that `the continuant that observes outcome $O_1$ observes outcome $O_1$,' which is hardly informative. Thus observations made by continuants like this do not suffice to confirm one theory over another, since we have no means of denoting the continuants which is independent of the outcomes of their observations. 

It has been argued that perhaps we can make sense of the individuation of Everettian continuants via indexical self-reference: for instance, Ismael claims that `I don't have to have a name or description that applies to a thing to refer to it indexically; I only have to be contextually related to it in the right way.' However, it is important to recognise the novelty of the kind of indexicality required by this view, for unlike ordinary indexical information, this indexical information cannot be given any analysis in terms of purely physical facts. Normally, indexical questions have a determinate answer because they are associated with particular physical thought-events: for example, Bostrom notes that there may be a quantum fluctuation somewhere in the universe which has produced a brain in a subjectively identical state to my own \cite{Bostrom}, but although I may be unable to answer the question `Am I the individual on planet Earth or the quantum fluctuation?' this question nonetheless has a determinate answer, because the thought must be associated with a physical thought-event at some spatiotemporal location, and the correct answer is determined by that location. \footnote{One might worry that if the two brains have such a thought at exactly the same time, we will not be able to say which thought belongs to which thought-event - after all, thoughts themselves don't appear to have physical locations, so unless there is some metaphyiscal link tying thoughts to the physical world, there would be no way to individuate the two simultaneous thoughts, let alone associate them with one physical thought-event rather than another. But the notion that there can be some question about which event a thought should be associated with presupposes that thoughts are ontologically distinct from physical thought-events, and therefore this sort of claim depends on an implicit dualism which is in tension with physicalism. Moreoever, in the context of special or general relativity, the claim that thoughts don't have physical locations seems less plausible. After all, thoughts do certainly seem to take place in time, and given that in relativistic theories we cannot assign a position in time without also assigning a position in space, it seems as though thoughts must indeed be associated with some spatial region, even if we are not generally aware of their physical locations.} And of course, certain kinds of Everettian indexical information may be understood similarly - for example, indexical information about one's position in the branching structure of an Everettian universe can be thought of as information about the branch in which the thought-event associated with some particular thought occurs. However, Ismael is claiming that there can be indexical individuation of pre-measurement continuants, and that sort of indexicality cannot be understood in the same way; as both Tappenden \cite{Tappenden} and Peter Lewis \cite{PeterLewis} have pointed out, a thought or utterance like `I wonder which outcome of the measurement I will see?' is associated with only a single physical event, and no information about the location and/or branch in which this event occurs singles out either of the future branches corresponding to the two continuants. Thus if there is indexical information involved here, it must be purely indexical in an entirely new way. It is, of course, possible that such novel indexical facts do exist, although this proposal sits uncomfortably with the physicalist approach, but we should recognise that the introduction of these facts is by no means a trivial extension of our ordinary understanding of indexicality. If we don't want to accept these sorts of facts, and we do wish to avoid the problem of ballooning credences, it seems that we must after all accept some belief-updating strategy which does uphold the Relevance-Limiting Thesis, and it follows that an Everettian version of quantum mechanics is not capable of empirical confirmation, so we have no reason to think that beliefs about quantum mechanics formed by reasoning from empirical evidence would still be true in an Everettian universe. 

\newpage

\section{Reference} 

Though the notion of truth is by no means a transparent one, it certainly seems reasonable to say that in order for our beliefs to be true, the concepts they invoke must successfully refer to real entities. Moreoever, it is a plausible condition on reference that a concept fails to refer to an actual entity if the resemblance between the concept and the entity is merely accidental. Putnam has a famous illustration: `An ant is crawling on a patch of sand ... by pure chance the line that it traces curves and recrosses itself in such a way that it ends up looking like a recognizable caricature of Winston Churchill ... similarity (of a certain very complicated sort) to the features of Winston Churchill is not sufficient to make something represent or refer to Churchill.' \cite{Putnam1981}. In order to deny this claim about  reference, we would have to argue that concepts can successfully refer to real entities even if those entities bear no responsibility for our having formed the concepts, which necessitates the adoption of a view like descriptivism (see Searle \cite{Searle1983}) on which the descriptive content associated with our concept of a theoretical entity is adequate to fix reference to a real entity. But descriptivism sits uncomfortably with the physicalism embraced by Oxford school Everettians, because descriptive theories of reference require us to attribute mental processes with an apparently ‘magical’ ability to attach to non-mental objects in the absence of any physical connection between brain and object (see Putnam 1981 \cite{Putnam1981} for further discussion of this objection), and therefore any Everettians who wish to remain faithful to physicalism should accept a non-accidental condition on reference.

However, it follows from such a condition that our beliefs about quantum mechanics can be true only if the theoretical entities invoked by quantum mechanics are in some sense responsible for our having formed concepts of those entities. Since we formed those concepts on the basis of empirical evidence, this means that the theoretical entities concerned must bear at least partial responsibility for our having made the observations that we have. In the ordinary probabilistic version of quantum mechanics, the theoretical entities are related to our evidence via the probabilities that they impart to various outcomes, and therefore the probabilities must in some sense be responsible for our having observed certain sequences of outcomes.  Similarly, in the Everettian case the theoretical entities are supposed to be related to our evidence via the mod-squared amplitudes that they assign to various outcomes, so the mod-squared amplitudes must in some sense be responsible for our having observed certain sequences of outcomes. This is problematic, because the Everettian theory claims that all possible sequences of observations are in fact made and does not award a privileged status to any particular observer. (see Hemmo and Pitowsky, 2007, \cite{Hemmo} and Albert, 2010 \cite{Albert2010}). Thus even if we do happen to occupy one of those branches in which the relative frequencies are close to the mod-squared amplitudes, this is purely a matter of good luck and not a fact for which the amplitudes bear any responsibility, so the amplitudes cannot possibly be responsible for our having made the observations that we have. As Albert \cite{Albert2010} puts it, the evidence for quantum mechanics `is of certain particular sorts of experiments having certain particular sorts of outcomes with certain particular sorts of frequencies - and not with others ... (Everettian quantum mechanics) is structurally incapable of explaining anything like that.' Moreover, the Everett approach entails not only that mod-squared amplitudes cannot play this role, but also that nothing else in the theory can play it either: since the theory does not single out any one sequence of outcomes, no entity defined within that theory can be responsible for our having seen the particular sequence of results that we have, and it follows that in an Everettian universe we would not be able to establish reference to the theoretical entities required by the Everettian theory. 

To put the problem in sharper focus, consider first the position of Everettian observers in those branches corresponding to sequences of outcomes in which the relative frequencies do not match the real mod-squared amplitudes. In most such branches observers will find it impossible to formulate any coherent theory at all, and if they do produce a theory it will probably not be even remotely similar to quantum mechanics. But suppose there is a branch in which the relative frequencies are anomalous in precisely the right way for the observers in that branch to formulate a quantum theory identical to ours except in that it assigns different numerical values to mod-squared amplitudes. When the observers in this branch speak about the theoretical entities postulated by their theory, they must fail to refer to actual entities, since the results which have led them to produce their theory are purely accidental. Now consider the observers in a branch where relative frequencies do in fact approximate the actual mod-squared amplitudes, and who therefore formulate a quantum theory identical to ours in all respects. There is no salient difference between these observers and the observers in the anomalous branch; none of the machinery of the theory has played any part in bringing it about that they observe this particular sequence of results rather than the sequence observed in the anomalous branch, so the correctness of their theory is still purely accidental and their concepts must still fail to make referential contact with actual entities. Thus no observer in an Everettian universe, whether or not they happen to occupy a branch in which the right relevant frequencies occur, would be able to refer successfully to the theoretical entities postulated by Everettian quantum mechanics.

It might be objected there is a sense in which the mod-squared amplitudes must be responsible for our observations, because everything that happens in an Everettian universe is emergent from the underlying universal quantum state of which the amplitudes are a part. But this is inadequate, because not just any kind of responsibility will suffice to fix reference. In Putnam's example, we might imagine that Churchill is somehow causally related to the ant's behaviour - perhaps the ant is crawling around on a bed of sand which Churchill laid down - but that kind of causal relationship is not enough to make the picture into a picture of Churchill. In order for the ant to be successfully referring, Churchill must have played a determining role in the ant's having formed the concept that it intends to depict; he must be causally responsible for its having formed that concept rather than another. To take an even closer parallel, if a scientist randomly forms a hypothesis about the brain which purely by coincidence bears some similarity to a real feature of the brain's operation, and if that feature just happens to play some role in the mental process by which he formulated the hypothesis, it doesn't follow that the concept automatically refers to the feature in question -  the establishment of reference requires that the feature is partially responsible for his having formed that particular concept rather than another. This is what causes problems for the Everett approach: mod-squared amplitudes cannot possibly be responsible for our having formed one quantum mechanical concept rather than another, since we formed our quantum mechanical concepts on the basis of a particular sequence of observations, and the Everettian version of quantum mechanics does not single out any one sequence of observations over the others. Thus we have exposed a fundamental incoherence in the Everett approach: it appears to follow from the theory itself that we cannot possibly succeed in forming true beliefs about the entities it invokes. 

\newpage

\section{Existing Approaches to the Probability Problem}

It has long been recognised that the Everett interpretation has a prima facie difficulty with probability and probabilistic confirmation, and a range of responses have been proposed \footnote{See Kent 1990 for a discussion of earlier approaches \cite{Kent}}. In this section, I briefly explain the difficulties with the intuitively appealing branch counting approach, which has in fact been rejected by the Oxford school Everettians themselves. I then turn to the two approaches which are currently in favour with the Oxford school - the 'no worse off' argument and the decision theoretic approach - and demonstrate that neither of these responses is adequate to address the incoherence in the Everett approach which has been identified in sections two and three. 

\subsection{Branch Counting} 
On first acquaintance with the Everett interpretation, it is tempting to think that probabilities in the Everett approach should be understood as probabilities for an observer to find himself in one branch rather than another after the measurement. The most intuitive way of filling in the numerical details, which Wallace \cite{Wallace2010Decoherence} calls `na\"{i}ve branch counting,' suggests that each of the branches should be regarded as equally likely. But it is easy to see that this gives the wrong probabilities, because it implies that when a measurement has two possible outcomes (such as `spin up' and `spin down,') the probability of each outcome must always be 0.5, whereas quantum mechanics tells us that the probability for an outcome is equal to the mod-squared amplitude for that outcome, which may be any number between 0 and 1. Moreoever, it is a central tenet of the Oxford school Everett interpretation that the branches are emergent features of the wavefunction rather than separate elements of the ontology, and therefore `there is no such thing as the number of branches ... (since) the branching structure is given by decoherence, and decoherence does not deliver a structure with a well-defined notion of branch count.' \cite{Wallace2010Decoherence} Thus in all but very simple cases the na\"{i}ve branch counting approach is untenable for Everettians of this school. 

A more sophisticated way of proceeding gets around these objections by postulating a continuous infinity of `worlds' and claiming that when a measurement is made, the fraction of these in which a certain outcome occurs is equal to the mod-squared amplitude for that outcome \footnote{A similar strategy is used in the many-minds approach, which postulates a continuous infinity of conscious minds; when a measurement is made, the fraction of these minds which observe a certain outcome is equal to the mod-squared amplitude for that outcome. See Papineau, 1995 \cite{Papineau1995}}. But if this strategy is invoked, it can no longer be claimed that the Everett approach does not require us to add anything to the basic quantum formalism, which is one of the cornerstones of the Oxford school Everett approach (see Wallace, 2001 \cite{Wallace2001}). More importantly, the move doesn't really put us in any better position with regard to probabilities; it introduces no uncertain events which could have probabilities other than zero or one, since it merely gives us some new certainties, that is, certainties about the proportion of worlds associated with each outcome. Hence we are still unable to make sense of nontrivial probabilities unless we postulate some stochastic process by which a single world is singled out as special - for example, by claiming that one world is somehow selected as the one to be experienced. But the plausibility of this idea rests on a metaphysical picture in which the mind is some kind of nonphysical entity that inhabits one branch or another, and we saw in section two that this view is incompatible with the metaphysical views underlying the Oxford school Everett approach. Moreoever, even if we avoid linking the stochastic mechanism with consciousness in this way, postulating such a mechanism is tantamount to abandoning the Everett picture in favour of a hidden variable interpretation, so this is not a viable means of defining genuinely Everettian probabilities.

\subsection{No Worse Off?} 

Since a satisfactory account of probability within the Everett interpretation has proven somewhat difficult to obtain, it has become popular for Oxford school Everettians to argue that other probabilistic theories are likewise incapable of giving an account of probability, and therefore the Everett approach is no worse off in this regard (see Papineau, 2010 \cite{Papineau2010}). And it is certainly true that even in non-branching cases, the introduction of probabilities makes it more difficult to understand how we come to form true beliefs about theoretical entities. When everything is deterministic, we can simply assert that the entities postulated by the theory are involved in underlying mechanisms that produce the phenomena on the basis of which we have formed our beliefs - for example, our theory of electrons tells us that excited electrons falling to lower energy levels cause the emission of light from cathode ray tubes, and therefore we can come to form true beliefs about electrons on the basis of experiments involving such tubes. But in a probabilistic theory the situation is more complex, since we cannot observe probabilities in any straightforward way. Consequently, once probabilities are introduced we require two steps to make sense of the relationship between our evidence and our beliefs: first, we assert that the entities postulated play some role in the underlying mechanisms that impart probabilities to various possible outcomes of our experiments, and second, we make a key assumption: 

\begin{assumption}
\label{Assumption1}
The relative frequencies of outcomes in the sequence of measurement outcomes we have observed are close to the probabilities that the underlying theory imparts to those outcomes. 
\end{assumption}
In the Everettian case, of course, rather than invoking probabilities, we assert that the entities postulated are involved in underlying mechanisms which assign mod-squared amplitudes to various possible outcomes of our experiments, and we make the revised assumption: 
\begin{assumption}
\label{Assumption1a}
The relative frequencies of outcomes in the sequence of measurement outcomes we have observed are close to the mod-squared amplitudes that the underlying theory assigns to those outcomes. 
\end{assumption}

The `no worse off' argument amounts to the claim that both Assumption \ref{Assumption1} and Assumption \ref{Assumption1a} are equally unjustified. This claim is given some traction by the fact that attempts to justify Assumption \ref{Assumption1} in terms of some particular account of the nature of probability, such as frequentism, have been found to be fraught with difficulties (see Hajek, \cite{Hajek1996}). However, it is possible to give a justification for Assumption \ref{Assumption1} without committing to any particular account of the nature of probabilities; we merely appeal to the `Principal Principle' originally formulated by Lewis \cite{Lewis1980}, which asserts that if an agent knows that the objective probability that E holds is p, then he is rationally compelled to set his credence in E to be p. It is common to say that that the only thing we really know about (objective) probability is that it must satisfy the Principal Principle; indeed, Wallace \cite{Wallace2006} claims that the Principle should be taken as a functional definition of probability. Thus we may argue: 

\begin{enumerate}
\item \textbf{The Principal Principle:} The rational credence to assign to the occurrence of outcome C, conditional on some proposition E which specifies an objective probability function P (and which contains no inadmissible information), is P(C).
\item The sequence of measurements S is independent and identically distributed. 
\item \textbf{The Law of Large Numbers:} As the length of a sequence of independent identically distributed measurements approaches infinity, the (objective) probability that the relative frequency of each outcome equals the (objective) probability of that outcome approaches one.
\end{enumerate}
\textbf{Conclusion:} I am rationally compelled to have high credence that the relative frequencies of outcomes in sequence S equal the objective probabilities of those outcomes (when S is sufficiently long). 
\vspace{2mm} 

Premise 1) is simply our functional definition of probability; premise 2) is a non-trivial empirical assumption, but we are entitled to make it because if the experimental setup which we are considering cannot be used to produce a series of this kind, the whole project of seeking to confirm a probabilistic theory about it is a non-starter; and premise 3) can be proven mathematically from the axioms of the probability calculus. 

If these three premises are accepted, we have a justification for Assumption \ref{Assumption1}, since the conclusion tells us that it is rational to make Assumption \ref{Assumption1} about any observed sequence of outcomes which is sufficiently long. Notice, however, that we are ultimately interested in forming beliefs that are not only rational but true, and thus this justification will be adequate only if the reason for the rationality is that having such credences is a good way of arriving at true beliefs. Since the rationality of making Assumption \ref{Assumption1} is supposed to be derived from the rationality of assigning credences as specified by the Principal Principle, we require that the reason it is rational to do this is that having such credences is a good way of arriving at true beliefs. If the reason for the rationality were of a very different nature - if, for example, it were grounded on decision-theoretic arguments similar to those examined in section 4 - the suggested justification for Assumption \ref{Assumption1} would not serve our purpose, since the rationality of making Assumption \ref{Assumption1} would then be unconnected to the question of whether Assumption \ref{Assumption1} is actually true. Indeed, we require not only that assigning credences as specified by the Principal Principle is a good way of arriving at true beliefs, but that this is so because of something to do with the nature of objective probabilities, since we saw in section three that in order for our concepts of theoretical entities to refer successfully in a probabilistic theory, the probabilities must in some sense be responsible for our having observed certain sequences of outcomes.

This means that the coherence of a probabilistic theory depends crucially on the claim that there is something in the world which not only satisfies the Principal Principle but which is causally responsible for the fact that credences formed in accordance with the Principal Principle tend to be true. However, it does not follow that we must be able to specify what this something actually is - in the absence of such a specification, probabilistic theories are arguably incomplete, but they are not incoherent - and therefore the rationality of believing in a probabilistic theory is not threatened by the fact that most such theories provide no substantive account of what it is in the world that satisfies the functional definition given by the Principal Principle. From an epistemological point of view the probabilities are really just placeholders: it is definitive of the theoretical substructures postulated by the theory that they impart certain probabilities to certain measurement outcomes, and therefore, given that what it is to be a probability is to satisfy the Principal Principle, the relationship between the theoretical entities and our observations is built directly into the theory

Can we give an analogue of this sort of justification in the Everettian case? Matters would be straightforward for the Everettian if we could simply replace `objective probability' with `mod-squared amplitude,' in the above argument, but this alteration is not trivial, because although we can appeal to the quantum analogue of the Law of Large Numbers \cite{Hartle} in order to obtain a quantum version of premise 3), we cannot so obviously make the same replacement in premise 1). We can certainly formulate a quantum version of the Principal Principle:

\vspace{2mm}
\emph{ \textbf{Quantum Principal Principle:} If I know the mod-squared amplitude corresponding to a measurement outcome to be p, then I am rationally compelled to set my personal credence in the occurrence of that outcome to be p.}
\vspace{2mm}

But Price \cite{Price} points out that there is an important distinction to be made between a probability map and the Born rule: `a probability map just is a practical guide to decision-making under uncertainty, so there's no mystery about why we use it that way,' whereas `the Born rule cannot be regarded the same way without some argument that the Everett picture allows an analogue of decision-theoretic uncertainty.' Although we are concerned here with the connection to truth rather than uncertainty, the same distinction is relevant: the Principal Principle is just part of what it means to be a probability, whereas the QPP cannot be accepted without some argument that amplitudes do actually perform the proposed function. In both cases, this distinction exists because amplitudes are physical quantities whose role in the theory is entirely separate from any role they might play in our epistemic practices: the quantum state, including the amplitudes attached to various branches, is given an unambiguous ontological significance in the Everett interpretation, so for example Wallace writes that the theory `read(s) the quantum state literally, as itself standing directly for a part of the ontology of the theory. To every different quantum state corresponds a different concrete way the world is.' \cite{Wallace2012} Since this ontology does not imply anything about the relationship between the amplitudes and our credences, the QPP cannot follow from it, and must therefore be a substantive claim which depends on empirical facts, not merely on the meaning of the concept of `mod-squared amplitude.' 

It might be objected that we could choose to make the QPP into a functional definition, and rather than defining mod-squared amplitudes in terms of their relation to the rest of the theory, define the postulated theoretical entities by their relation to mod-squared amplitudes: then we can argue, as in the probabilistic case, that it is definitive of the theoretical substructures postulated by Everettian quantum mechanics that they impart certain mod-squared amplitudes to certain measurement outcomes, and therefore, given that what it is to be a mod-squared amplitude is to satisfy the QPP, the relationship between the theoretical entities and our observations is built directly into the theory. But this is not viable, because neither the QPP nor the Everettian theory singles out any one of the many sets of actually occurring sequences of measurement outcomes, so the relation between the entities in question and our particular history of observations cannot possibly follow from just the concepts of the theory together with the QPP. In order to secure such a relation, we would need to alter the functional definition of amplitudes so that it implies that they are related in the way specified by Assumption \ref{Assumption1a} to our own particular observations, regardless of what other observations are made in other branches; but this would force our fundamental ontology to have an irreducibly indexical component, and thus just as in section 2.2.1 we would be forced to postulate an entirely new sort of indexical information. If we are unwilling to do so, the justification of Assumption \ref{Assumption1} by appeal to the meaning of probability cannot be translated to an analogous justification of Assumption \ref{Assumption1a} in the Everettian case, and therefore the Everett interpretation is genuinely worse off than other probabilistic theories in this regard.

\subsection{Decision Theory}

The approach to probability presently favoured by the Oxford school Everettians involves using the methods of decision theory to prove that the only rational way for an Everettian agent to assign credences to measurement outcomes is in accordance with the Born rule (see Deutsch \cite{Deutsch1999}, Wallace \cite{Wallace2010}, Greaves \cite{Greaves2007}). Many existing critques of the decision theoretic approach focus on specific details of the proofs - for example, Dizadji-Bahmani focuses on the justification of the Branching Indifference axiom \cite{DizadjiBahmani}, while Baker objects to the role played by decoherence \cite{Baker}. However, in this section I will consider the broader question of whether decision theory can possibly be the right sort of tool to address the issue of confirmation in the Everett interpretation. I will argue that probabilities understood in a decision-theoretic sense are incapable of playing the required role in reasoning leading to true beliefs, and therefore decision theory is not an adequate way of dealing with the Everettian epistemic problem. 

\subsubsection{Decision Theory in the Everett Interpretation}

In classical decision theory, we start by assuming that agents have a set of preferences over possible actions which obey certain general principles of rationality. We then prove that these preferences can be (almost) uniquely represented by a credence function and a utility function over possible outcomes such that the agent always chooses the action which maximises expected utility, where the expected utility for action $A$ is equal to $ \sum_i C(o_i | A) U(o_i)$, with $C(o_i | A)$ giving the credence attached to outcome $o_i$ conditional on $A$ being carried out, and $U(o_i)$ giving the utility attached to outcome $o_i$ (see Wallace, 2012 \cite{Wallace2012}). A similar strategy may be applied in the Everettian context; again assuming that the agent has a set of preferences over possible actions which obey certain general principles of rationality, it can be shown that the the agent's preferences may be represented by a quasi-credence function and a utility function over branches of the wavefunction such that the agent always chooses the action which maximises expected utility calculated as in the classical case, and moreoever, that the quasi-credences must be equal to the mod-squared amplitudes for the branches. Hence, it is claimed, a rational agent in an Everettian universe will act just like a rational agent in a classical universe who assigns credences which equal mod-squared amplitudes.

The original decision-theoretic proofs given by Everettians dealt only with decision-making within an Everettian universe on the part of an agent who is sure that Everettian quantum mechanics is true and knows the quantum state of the universe. But the approach has been generalised to the case of an agent who is deliberating over a range of hypotheses, including possibly Everettian quantum mechanics or other branching theories, and/or does not know the quantum state of the universe (see Greaves, 2007 \cite{Greaves2007}); Wallace gives a more comprehensive representation theorem, the Everettian Epistemic Theorem, which tells us that the preferences of such an agent are represented by utility and quasi-credence functions over both possible worlds and branches of the wavefunction, with the quasi-credence function conditional on Everettian quantum mechanics being true given by some density operator which is updated by unitary dynamics and Bayesian conditionalisation \cite{Wallace2012}. If the question of how it is rational to update our beliefs can correctly be regarded as just a question about how it is rational to act, it follows that an agent considering a range of theories including Everettian quantum mechanics should update his beliefs just as we ordinarily do, treating mod-squared amplitudes as we ordinarily treat probabilities - that is, he should apply Extended Conditionalization. Since applying this method of conditionalization leads to a credence distributions over centered worlds which is numerically identical to the credence distribution that follows from ordinary Bayesian updating over the corresponding possible worlds in the non-branching case, these proofs appear to show that it is rational for us to reason about Everettian hypotheses with probabilities replaced by mod-squared amplitudes just as we usually reason about the probabilistic versions, which suggests that Everettian version of quantum mechanics has exactly the same degree of confirmation as ordinary probabilistic quantum mechanics. 

\subsubsection{ Rationalizing Decision Theory}

However, these putative proofs of the Born rule are acceptable only if we first accept that decision theory is a valid approach to defining Everettian probabilities. Proponents of the decision-theoretic approach are certainly aware that some justification is required for this; in particular, many attempts have been made to explain why decision theory, which is supposed to be a theory of rational decision-making under uncertainty, can still be applied in a context where every measurement outcome always occurs so there is no ordinary uncertainty (see Hemmo and Pitowsky, 2007 \cite{Hemmo}). Two ways of addressing this difficulty have been pioneered: the `subjective uncertainty' viewpoint claims that the mental state of an Everettian agent approaching a branching event must be identical to the mental state of an agent approaching an ordinary uncertain event, and therefore his situation is functionally identical to a state of classical uncertainty and his quasi-credences should come out just as his credences would in the classical case (see Saunders 2005 \cite{Saunders2005}), while the `objective deterministic' viewpoint accepts the absence of uncertainty at branching events, and interprets quasi-credences over measurement outcomes as measures of how much the agent cares about the successor in each branch (see Greaves 2004 \cite{Greaves2004}). 

Both of these approaches have been subject to extensive criticism, and it is by no means obvious that either of them succeeds in locating a genuine source of uncertainty within the Everett picture (see Peter Lewis 2006 \cite{PeterLewis} and Greaves 2004 \cite{Greaves2004}). Moreoever, I claim that this focus on uncertainty has obscured the true locus of the difficulty with the application of decision-theoretic methods in the Everettian context, which is the conceptual link between decision-theoretic belief and truth. In brief, decision theory tells us how it is rational to act; thus these decision-theoretic proofs, if they succeed, can at best demonstrate that it is rational to act as if we assign credences according to the Born rule. But the fact that it is rational to act as if we have certain beliefs does not imply that it is rational to actually hold those beliefs, unless we accept some variant of behaviourism, which is by no means an uncontroversial philosophical position. \footnote{ Behaviourism is the claim that to hold a belief just is to exhibit all the behaviours associated with this belief. Variants of behaviourism have been espoused by writers like Dennett \cite{Dennett} and Davidson \cite{Davidson}, although it also has influential opponents, including Searle \cite{Searle}, Fodor \cite{Fodor} and Kim \cite{Kim1980} } Furthermore, even if we accept behaviourism, the proofs show only that it is rational to update our beliefs via Extended Conditionalization; they do nothing to convince us that the reason for the rationality is that Extended Conditionalization is a good way of forming true beliefs. This is problematic if decision theory is to be applied to the question of the empirical confirmation of the Everettian theory, since in scientific theory confirmation our concern is to find out what is true of the world, and therefore rational belief matters only instrumentally as a means of getting at truth.

In what follows, I demonstrate that the Everett interpretation is a context in which the usual conceptual connections between rational belief and truth fails to hold, and therefore decision theory is not an appropriate way of addressing the epistemic problems I raised in sections two and three. First, I explain several ways in which branches are relevantly different to possible worlds. Second, I show that the Everettian hypothesis is an example of a wider class of hypotheses in which we are related to our evidence in a seriously deviant way, and which consequently must be excluded from consideration when we make inferences based on empirical evidence. Finally, I discuss a proposal by Greaves and Wallace which involves using epistemic rather than pragmatic utilities in order to restore the connection between decision-theoretic belief and truth; I argue that this proposal does not succeed, because the use of such epistemic utilties undermines important presuppositions of the decision-theoretic approach.

\paragraph{Branches versus Worlds}

The use of decision-theory in the Everettian context requires us to assume that an agent's preferences are represented by quasi-credences and utilities attached to branches of the wavefunction rather than possible worlds - that is, branches are to be treated in our reasoning processes just as we normally treat possible worlds. But this is questionable on a conceptual level, since it obliterates the distinction between branches and worlds: there is nothing in the mathematics which distinguishes the collection of branches which together constitute a possible world from any arbitrary collection of branches, and consequently the decision-theoretic approach requires that we reason as if the problem of locating oneself within a world is exactly equivalent to the problem of deciding which world one is in, whereas at least intuitively it seems that these are very different sorts of epistemic problems.

For example, Price points out that we have no good reason to assume an Everettian agent's preferences will be attached to individual branches at all, because the existence of branching structure provides `a new locus for possible preferences' (\cite{Price}) - that is, the agent may have preferences about the overall branching structure of the universe, preferences of a sort which would make no sense in a classical context. Moreoever, Price asserts that credences have normative force because of what he calls the `credence-existence' link: credences reflect degrees of confidence about whether things actually exist, and we only care about things which actually exist, so we are justified in disregarding possibilities to which we assign low credence. In the classical context, the decision-theoretic approach requires agents to ignore to possible worlds which diverge from the actual world before the present time, and we can rationalize this in terms of the credence-existence link by noting that unless we are Lewisian realists, other possible worlds are not physical objects which exist and evolve independently of us \footnote{ Even for Lewisian realists they are not part of the actual world, which is presumably what we are aiming to describe with our scientific theories}, so we have zero confidence in the existence of worlds which diverge from our own before the present, and are therefore entitled to ignore them. Whereas in the Everettian context, treating branches like possible worlds requires us to ignore branches which are connected to our own branch in the past but not the future; but such branches are still out there in the universe, physically real and evolving independently of us, and therefore we do not have the same kind of justification for ignoring them \footnote{Wilson claims that we are free to choose between two different metaphysical pictures of the Everettian theory: either we treat each Everettian multiverse as a Lewisian possible world (Collectivism) or we treat each distinct branch in an an Everettian multiverse as a Lewisian possible world (Individualism), and if we choose the latter then we are justified in ignoring Everettian branches connected to our own in the past but not the future. However, whatever metaphysical picture we adopt, recall that quantum mechanics purports to describe the entire universe including all the branches of the wavefunction, and therefore the belief-updating strategy we apply to deliberations over the truth of quantum mechanics should not advocate ignoring most of those branches. Moreoever, unlike  Lewisian possible worlds, other Everettian branches could at least in principle interact with our own branch and cause interference effects, so it seems implausible to say that the choice about whether or not we may ignore them is merely a matter of metaphysics (in practice the probability of enough coherence being achieved is vanishingly small, but this sort of practical point does not seem adequate justification for behaving as if other branches are somehow less real). }. Thus rational belief-updating based on preferences attached to individual branches considerations will fail to give us access to wider truths about the full branching structure, so decision-theoretic methods based on in-branch preferences are not an adequate way of understanding how we should attempt to form true beliefs in such a context.  

Likewise, if in the Everettian context we treat branches like possible worlds, the quasi-credence attached to branch B within branching world W should be equal to the product of the credence attached to world W and the quasi-credence attached to B conditional on W (which is given by the mod-squared amplitude assigned to B by the theory governing world W); this composition of a credence and a quasi-credence is then treated in just the same way as the total credence attached to a non-branching world. Thus if we are to rationalize the application of decision theory in this context, it must make sense to think of quasi-credences as the same kind of thing as ordinary credences - not just mathematically, but also conceptually. Yet neither of the usual Everettian approaches to understanding uncertainty enables us to think this way. If we accept the Subjective Uncertainty viewpoint, the composition of credences and quasi-credences is equivalent to the composition of real uncertainty with what is only a subjective impression of uncertainty: there is a fact of the matter about which underlying theory of the universe is true, and the agent's credences over possible worlds quantity his genuine uncertainty about that fact, but there is no fact of the matter about which measurement outcome he will see, since every outcome is seen by one of his successors, so his quasi-credences over outcomes quantify only an impression of uncertainty. Whereas if we accept the Objective Deterministic viewpoint, the composition of credences and quasi-credences is equivalent to the composition of ordinary credences with a measure of how much we care about various future people; moreoever, the ordinary credences we assign to various possible scientific theories cannot be interpreted as measures of how much we care about our counterparts in the possible worlds governed by those theories, because allowing considerations about how much one cares about particular individuals to impinge on credences (as opposed to quasi-credences) is a paradigmatic example of epistemic irrationality. Thus it does not seem as though we have any good justification for treating quasi-credences in the way that the decision theoretic approach requries, and thus belief-updating based on treating branches like worlds cannot be an adequate model for epistemic rationality in an Everettian universe, because it fails to take into account the important epistemic consequences of the fact that the branches are indeed branches and not worlds.

\paragraph{Deviant hypotheses} 

In sections two and three, we saw that the epistemology of the Everett interpretation is rendered problematic by the fact that the Everettian hypothesis implies that we are related to our evidence in a way that differs significantly from our usual conception of that relationship. The Everett interpretation thus belongs to a wider class of hypotheses which propose a severely deviant relationship between evidence and observers. In the context of such hypotheses, the usual connections between rational behaviour, belief and truth are undermined: for  example, I believe that I may be a brain in a vat, though this is not a possibility to which I normally pay much attention, nor one which influences my behaviour to any significant degree (except when I am writing about philosophy!) and a decision-theoretic analysis would therefore yield the conclusion that my credence in the possibility must be low. Yet in fact, the reason I don't act on this hypothesis isn't that I think it is probably false; I simply have no idea how likely it is to be true, and I am in any case unsure how I ought to respond if it were true. Thus beliefs about hypotheses like this are simply not linked to rational behaviour in any systematic way, and therefore the usual connection between belief that is rational in a decision-theoretic sense and truth does not hold in this context. 

Furthermore, the difficulties that arise once we begin to consider such hypotheses are not limited to beliefs in the hypotheses themselves. Normally, when a coin is flipped, I would say - and a decision-theoretic analysis of my behaviour would no doubt confirm - that I have a credence of about 50\% in the possibility that the coin will land heads, and about 50\% in the possibility that the coin will land tails. But the coin will not land either heads or tails if I am a brain in a vat, so the sum of these two credences should be less than 100\% by an amount corresponding to the credence I attach to the possibility that I am a brain in a vat. Since my credence in that possibility can't be quantified, it seems as though my credence in heads or tails won't be quantifiable either, and similar reasoning will apply to more or less any empirical belief that I might have, and therefore allowing such hypotheses undermines the connection between rational behaviour and truth even in the case of beliefs which are not obviously related to the hypotheses. 

Indeed, when such radical possibilities are under consideration, we have good reason to question an important presupposition of  the decision theoretic approach - namely, that an agent will have a set of preferences over possible actions at all. In general, people form preferences over possible actions based on their judgements about what future experiences those actions are likely to lead to, so it is quite reasonable to suppose that an agent might simply be unable to form any preferences over actions when it is not clear how to make judgements about the likelihood of future experiences: for example, it may be the case that the epistemically rational thing to do when faced with a branching-universe hypothesis is to simply disregard that possibility, and if this is correct, it follows that that rational agents ought to refrain from forming any preferences over actions conditional on the truth of a branching-world hypothesis.  \footnote{Greaves herself raises the possibility of this sort of nihilism in an Everettian context (see Greaves, 2004 \cite{Greaves2004}) and Price also aruges for its plausibility.} Thus the basic starting point of the decision theoretic approach may well be illegitimate when we are contemplating hypotheses such as Everettian quantum mechanics, and therefore decision theory is not capable of giving us an understanding of the right way for an Everettian agent to form beliefs that are true rather than just rational in the decision-theoretic sense.

\paragraph{Epistemic Utilities}

 In order to restore the connection between rational belief and truth in the decision-theoretic context, Greaves and Wallace discuss the possibility of replacing ordinary utilities with epistemic utilities, assigning to each possible pairing of a state of the world $S$ and a belief state $B$ a number which is supposed to represent the epistemic value of being in $B$ when the true state of the world is $S$ (see Greaves and Wallace, 2006 \cite{GreavesWallace}). However, there are difficulties with this approach even in the ordinary non-branching case: if we accept that epistemic utilities are distinct from pragmatic utilities, and that rational action may be influenced by both sorts of utilities, our decision theory will have a problem with underdetermination, because a description of an agent's behaviour together with a complete specification of his two utility functions will no longer serve to fix his credences, as it cannot settle the question of how epistemic and pragmatic utilities are weighted in any particular case. Greaves and Wallace \cite{GreavesWallace} foresee this problem and choose to limit the application of epistemic utilities to cases where the action in question is a decision about what to believe. But in real situations pragmatic utilities inevitably impose on decision problems of this nature, so unless there is some uniquely rational way to decide the relative weighting of epistemic and pragmatic utilities in any particular decision problem, there will be no unique probabilities that the agent ought to assign, and therefore there will no longer be any uniquely rational beliefs that an agent ought to hold, meaning that decision-theoretic rationality will not be suitable as a guide to truth. 

Could there in fact be a uniquely rational way to weight epistemic and pragmatic utilities? It does not seem plausible to hold that the weighting is always the same, because if believing truths were always of importance equal to our other pragmatic goals, then rational agents would always be mandated to go out and actively search for truths, and someone who chose to sit around and do nothing rather than spend his time carrying out experiments would qualify as irrational rather than merely lazy. So variations in weighting may occur, and deciding the appropriate weightings in any given case is itself a decision problem, which means we must either define a set of super-utilities which can be used to evaluate the relative merits of different possible weightings, or accept that there are some decision-making processes which do not fit under the decision-theoretic framework. The former is implausible, because the point of the utility function is that together with the probability function it fully determines the agent's decision-making behaviour, whereas these super-utilities would be sufficiently abstract that it would be difficult to establish the required kind of behavioural link. But the latter is counterproductive to the decision-theoretic approach, because it prevents us from claiming that all decision processes should be describable within the same overarching framework, which undercuts the motivation for attempting to subsume deliberations over belief under the decision-theoretic framework. Once we acknowledge that at least some decision processes cannot be understood within this framework, the fact that we recognise a distinction between epistemic and pragmatic motivations strongly suggests that epistemic deliberations may be another sort of process which should not be understood in this way, and thus we have good reason to think the conceptual connection between decision-theoretic rationality and truth may fail to hold in unusual contexts such as the Everett interpretation.

\newpage

\newpage

\section{Conclusion}

The plausibility of the Everett interpretation depends crucially on the claim that the empirical confirmation attached to our existing quantum formalism can be transferred to Everettian quantum mechanics. I have argued that this transfer is illegitimate, because it is not possible to claim that the kinds of reasoning we have used to arrive at our beliefs about quantum mechanics would still lead us to form true beliefs in an Everettian universe: in such a universe, quantum mechanics would not be capable of empirical confirmation on the basis of the kinds of evidence that we have for it, and in fact, we would not even be able to establish reference to the theoretical entities it invokes.

I have also shown that three particularly prominent approaches to Everettian probability are inadequate to address this difficulty. In particular, I have argued that the decision-theoretic approach currently favoured by the Oxford school cannot be applied to questions of theory confirmation because the connection between rational belief and truth is not maintained in this context. Although a number of rationalizations for this approach are available, they sidestep the most important conceptual problems by focusing attention on uncertainty rather than on the connection between rational belief and truth.

There are certainly ways in which the Everettian might respond to the arguments I have presented, but the response requires the adoption of a very specific set of philosophical theses, and there is something problematic about a scientific theory which is only coherent provided that we accept a certain set of largely semantic claims. Moreover, the philosophical package that emerges is unstable as a whole. For example, we saw in section 4.3.2 that in order to mount the decision-theoretic defence of Everettian probability, it is necessary to espouse behaviourism about belief so that it can be argued that the rationality of holding certain beliefs follows from the rationality of acting as if we hold those beliefs. But behaviourism involves an eschewal of talk about mental processes and entities, and is therefore inconsistent with the descriptivist thesis that concepts come to be meaningful via the mental processes by which we associate content with them; it is no coincidencethat  Searle, one of the most prominent defenders of descriptivism \cite{Searle1983} is also one of the most prominent critics of behaviourism \cite{Searle}. And we saw in section 3 that Everettians would need to make precisely such a descriptivist claim in order to maintain that the theoretical concepts of Everettian quantum mechanics can successsfully refer to actual entities even when the resemblance between the concepts and the entities is merely accidental. In a similar vein, behviourism is generally linked with a broadly physicalist background, and accordingly, most Oxford school Everettians identify themselves as physicalists; but we saw in sections 2.2.1 and 4.2 that two distinct arguments in support of the Oxford school Everett approach require the postulation of irreducibly indexical facts, which sit uncomfortably with strict physicalism. Thus there is no unified worldview which enables a complete defence of the Oxford school Everett interpretation to be made, and in the absence of any consistent defence the approach must be taken as internally incoherent.

\newpage 

\Large

\textbf{Acknowledgements}

\normalsize

\vspace{4mm} 
I would like to thank my thesis supervisor Dr David Wallace both for his encouragement and for his insightful criticisms, which were in invaluable in shaping and refining the ideas put forward in this article.


\begin{thebibliography}{9}

\bibitem{Albert2003}
D. Albert. `Time and Chance.' (Harvard University Press, 2003)



\bibitem{Albert2010}
D. Albert. `Probability in the Everett Picture'. In S. Saunders, J. Barrett, A. Kent and D. Wallace (eds.), \emph{Many Worlds? Everett, Quantum Theory, and Reality'} (Oxford University Press, 2010)


\bibitem{AlbertLoewer}

D. Albert and B. Loewer. `Interpreting the Many-Worlds Interpretation.' \emph{Synthese} Vol. 77, 195 - 213 (1988)

\bibitem{Baker} 
D. Baker. `Measurement Outcomes and Probability in Everettian Quantum Mechanics.' (2006)

\bibitem{Barrett}
J. Barrett. `Empirical Adequacy and the Availability of Reliable Records in Quantum Mechanics.' \emph{Philosophy of Science} Vol. 63, 49 - 64 (1996) 

\bibitem{Bonjour}
L. Bonjour. `Externalist Theories of Empirical Knowledge.' \emph{Midwest Studies in Philosophy} Vol. 5 (1):53-73 (1980)

\bibitem{Bostrom}
N. Bostrom. \emph{Anthropic Bias}. (Routledge, 2002)

\bibitem{Bradley}
D. Bradley. `Four Problems about self-locating Belief.' Philosophical Review Vol. 121, Number 2: 149 -177 (2012)

\bibitem{Carnap}
R. Carnap. \emph{Meaning and Necessity} (Chicago University Press, 1947) 

\bibitem{Davidson} 
D. Davidson. `Belief and the basis of meaning.' \emph{Synthese} Vol. 27, 309 - 323 (1974)

\bibitem{Dennett}
D. Dennett. `True Believers: The Intentional Strategy and Why it Works'. In A. Heath (ed.) \emph{Scientific Explanation} (Oxford University Press, 1981)

\bibitem{Dennett1987}
D. Dennett. \emph{The Intentional Stance}. (MIT Press, 1987) 

\bibitem{Deutsch1996}
D. Deutsch. `Comment on Lockwood'. \emph{British Journal for the Philosophy of Science} Vol. 47, 222228, (1996)

\bibitem{Deutsch1999}
D. Deutsch. `Quantum Theory of Probability and Decisions'. \emph {Proceedings of the Royal Society of London} A455, 3129 -3137E (1999)

\bibitem{DeWitt} 
B. DeWitt. `The Many-Universes Interpretation of Quantum Mechanics.'  In B. Espagnat (ed.) \emph{Foundations of Quantum Mechanics} (New York: Academic Press, 1971)

\bibitem{DizadjiBahmani}
Dizadji-Bahmani. `The Probability Problem in Everettian Quantum Mechanics Persists.' \emph{British Journal of Philosophy of Science} (2013) 

\bibitem{Dummett1993}
M. Dummett. \emph{Seas of Language} (Oxford University Press, 1993) 

\bibitem{Elga}
A. Elga. self-locating belief and the Sleeping Beauty problem. Analysis Vol. 60 (2) (2000)

\bibitem{Everett1957}
H. Everett. `Relative State’ Formulation of Quantum Mechanics.' \emph{Reviews of Modern Physics} Vol, 29, 454 - 462 (1957)

\bibitem{Everett}
H. Everett. In B. DeWitt. and R. Graham (eds.) \emph{The Many-Worlds Interpretation of Quantum Mechanics, Princeton Series in Physics} (Princeton University Press, 1973)

\bibitem{FarhiGoldstoneGutmann}
Farhi. J. Goldstone and S. Gutmann. \emph {Ann. Phys.} Vol. 192, 368 (1989)

\bibitem{Fodor}

J. Fodor. `Language, Thought and Compositionality.' 
\emph{Mind and Language} Vol. 16 (1), 1-15 (2001)

\bibitem{GellmannHartle}
M. Gellmann and J. Hartle. `Quantum Mechanics in the Light of Quantum Cosmology.' In W. H. Zurek (ed.) \emph{Complexity, Entropy, and the Physics of Information}, Proceedings of the Santa Fe Institute Studies in the Sciences of Complexity, Volume 8, 425 - 458 (Addison-Wesley, 1990)

\bibitem{Ghirardi}
G. Ghirardi. `Collapse Theories.' \emph{The Stanford Encyclopedia of Philosophy} (Winter 2011 Edition), Edward N. Zalta (ed.) 

\bibitem{Greaves2004}
H. Greaves. `Understanding Deutsch's probability in a deterministic multiverse.' \emph{Studies in History and Philosophy of Modern Physics} Vol. 35 (3) (September 2004)

\bibitem{Greaves2007}
H. Greaves. `On the Everettian epistemic problem'. \emph{Studies in History and Philosophy of Modern Physics} Vol. 38 (1), (March 2007)

\bibitem{GreavesWallace}
H. Greaves and D. Wallace. `Justifying conditionalization: Conditionalization maximizes expected epistemic utility'. \emph{Mind} Vol. 115, 459, (July 2006)

\bibitem{FarhiGoldstoneGutmann}
Farhi. J. Goldstone and S. Gutmann. \emph {Ann. Phys.} Vol. 192, 368 (1989)

\bibitem{GrinsteadSnell}
C. Grinstead and J. Snell. \emph{Introduction to Probability} (American Mathematical Society, 1998)

\bibitem{HalpernTuttle}
J. Halpern and M. Tuttle. `Knowledge, Probability and Adversaries.' \emph{Journal of the ACM} (1993)

\bibitem{Halpern}
J. Halpern. `Sleeping Beauty Reconsidered: Conditioning and Reflection in Asynchronous Systems.' \emph{Ninth International Conference on Principles of Knowledge Representation and Reasoning} (2004)

\bibitem{Hajek1996}
A. H\'{a}jek. `Mises Redux' - Redux: Fifteen Arguments against Finite Frequentism'. \emph{Erkenntnis}, Vol. 45 2/3 `Probability, Dynamics and Causality' (November 1996)

\bibitem{Hajek2009}
A. H\'{a}jek. `Fifteen Arguments Against Hypothetical Frequentism'. \emph{Erkenntnis}, Vol. 70 2 (March 2009)

\bibitem{Hartle}
J. Hartle. 'Quantum Mechanics of Individual Systems.' \emph{American Journal of Physics}, Vol. 36, 704 (1968)

\bibitem{Hemmo} 
M. Hemmo and I. Pitowsky. `Quantum probability and many worlds.' \emph {Studies in History and Philosophy of Modern Physics} Vol. 38, 333 – 350 (2007) 

\bibitem{HowsonUrbach}
C. Howson and P. Urbach. \emph{Scientific Reasoning: The Bayesian Approach}. Second Edition (Open Court Publishing Company, 1993)

\bibitem{Ismael} 
J. Ismael. `How to combine chance and determinism: Thinking about the future in an Everettian universe.' 	\emph{Philosophy of Science} (October 2003) 

\bibitem{Kaplan}
M. Kaplan. \emph{Decision Theory as Philosophy} (Cambridge University Press, 1998)

\bibitem{Kent} 
A. Kent. `Against Many-Worlds Interpretations.' \emph{International Journal of Modern Physics A}, 1745 (1990) 

\bibitem{Kim}
J. Kim. 'Events as Property Exemplifications' in M. Brand and D. Walton (eds.), \emph{Action Theory} (Reidel, 1976) 

\bibitem{Kim1980}
J. Kim. `Reasons and the First Person.' In J. Kim \emph{Essays in the Metaphysics of Mind} (Oxford University Press, 2010)

\bibitem{Kripke}
S. Kripke. `Naming and Necessity'. In D. Davidson and G. Harman (eds.), \emph{Semantics of Natural Language}, (Reidel, 1972) 

\bibitem{Krips} 
H. Krips. `Measurement in Quantum Theory.' \emph{The Stanford Encyclopedia of Philosophy} (Fall 2013 Edition), Edward N. Zalta (ed.)

\bibitem{LeitgebPettigrew}
H. Leitgeb and R. Pettigrew. `Bayesianism II: the Consequences of Minimizing Inaccuracy.' \emph{Philosophy of Science} Vol. 77 (April 2010) 

\bibitem{Leslie}
J. Leslie. `Observer-relative Chances and the Doomsday argument.' \emph{Inquiry} Vol. 40, 427-36 (1997)

\bibitem{Lewis1973}
D. Lewis. `Radical Interpretation.' \emph{Synthese} Vol. 27, 331-344 (1974)

\bibitem{Lewis}
D. Lewis. `Attitudes De Dicto and De Se.' \emph{The Philosophical Review} Vol. 88: 513 -543 (1979)

\bibitem{Lewis1980}
D. Lewis. `A Subjectivist's Guide to Objective Chance.' In R. Jeffrey (ed.), \emph{Studies in Inductive Logic and Probability} (University of California Press, 1980)

\bibitem{PeterLewis}
P. Lewis. `Uncertainty and Probability for Branching Selves.' (2006) 

\bibitem{Meacham}
C. Meacham. `Sleeping Beauty and the Dynamics of De Se Belief.' Forthcoming in \emph{Philosophical Studies}

\bibitem{Nola}
R. Nola. `Fixing the Reference of Theoretical Terms'. \emph{Philosophy of Science}, Vol. 47 4, 505 -531 (1980)

\bibitem{Papineau1995}
D. Papineau. `Probabilities and the Many Minds Interpretation of Quantum Mechanics'. \emph{Analysis}, Vol. 55 4 (1995)

\bibitem{Papineau1996Intro}
D. Papineau. `Introduction'. In D. Papineau (ed.) \emph{The Philosophy of Science} (Oxford University Press, 1996) 

\bibitem{Papineau1996}
D. Papineau. `Many Minds Are No Worse Than One'. \emph{British Journal for the Philosophy of Science} Vol. 47, 234 -41 (1996)

\bibitem{Papineau2010}
D. Papineau. `A Fair Deal for Everettians.' In S. Saunders, J. Barrett, A. Kent and D. Wallace (eds.), \emph{Many Worlds? Everett, Quantum Theory, and Reality} (Oxford University Press, 2010)


\bibitem{Parker}
D. Parker. `Thermodynamic Irreversibility: Does the Big Bang Explain What It Purports to Explain?' \emph{Philosophy of Science} Vol. 72 5, 751 - 757 (2005) 

\bibitem{Price}
H. Price. `Decisions, Decisions, Decisions: Can Savage Salvage Everettian Probability?' (2008) 

\bibitem{Putnam1963}
H. Putnam. `Brains and Behavior.' In Ronald J. Butler (ed.), \emph{Analytical Philosophy: Second Series} (Blackwell, 1963)

\bibitem{Putnam1975}
H. Putnam. \emph{Mind, Language and Reality: Philosophical Papers: Vol. 2} (Cambridge University Press, 1975)

\bibitem{Putnam1978} 
H. Putnam. \emph{Meaning and the Moral Sciences} (Routledge and Kegan Paul, 1978)

\bibitem{Putnam1981}
H. Putnam. \emph{Reason, Truth and History} (Cambridge University Press, 1981)

\bibitem{Quine1960}
W. Quine. \emph{Word and Object} (MIT Press, 1960) 

\bibitem{Salmon}
W. Salmon. \emph{The Foundations of Scientific Inference}. (University of Pittsburgh Press, 1996)

\bibitem{Saunders2005}
S.Saunders. `What is Probability?' In A. Elitzur, S. Dolev and N. Kolenda (eds.), \emph{Quo Vadis Quantum Mechanics} (Springer Verlag, 2005)

\bibitem{SaundersWallace}
S. Saunders and D. Wallace. `Branching and Uncertainty.' \emph{British Journal for the Philosophy of Science} Vol. 59 (2008)

\bibitem{Searle}
J. Searle. `Minds, brains, and programs.' \emph{ Behavioral and Brain Sciences} Vol. 3, 417-457 (1980)

\bibitem{Searle1983}
J. Searle. \emph{Intentionality} (Cambridge University Press, 1983)

\bibitem{Tappenden} 
P. Tappenden. `Saunders and Wallace on Everett and Lewis.' (2008) 

\bibitem{Tegmark}
M. Tegmark. `The Interpretation of Quantum Mechanics: Many Worlds or Many Words?' \emph{Fortschritte der Physik} Vol 46, 855-862 (1998)

\bibitem{Titelbaum}
M. Titelbaum. `The Relevance of self-locating Beliefs.' \emph{Philosophical Review} Vol. 117 4 (2008)

\bibitem{VanFraassen1984}
B. Van Fraassen.`Belief and the Will.' \emph{The Journal of Philosophy} Vol 81, 235 - 56 (1984) 

\bibitem{Vaidman} 
L. Vaidman. `Probabilities in the Many-Worlds Interpretation of Quantum Mechanics.' In Y. Ben-Menahem and E. Hemmo (eds.), \emph{Probability in Physics} (Springer, 2012)

\bibitem{Votsis}
I. Votsis. `Historical Case Study: The Caloric Theory of Heat.' In PhD thesis (London School of Economics, 2004) 

\bibitem{Wallace2001} 
D. Wallace. `Worlds in the Everett Interpretation.' \emph{Studies in the History and Philosophy of Modern Physics} Vol. 33, 637 – 661 (2002)


\bibitem{Wallace2003}
D. Wallace. `Everett and Structure.' \emph{Studies in the History and Philosophy of Modern Physics} Vol. 34, 86 - 105 (2003)

\bibitem{Wallace2006}
D. Wallace. `Epistemology Quantised: circumstances in which we should come to believe in the Everett interpretation.' \emph{British Journal for the Philosophy of Science} Vol. 57, 655 - 689 (2006)

\bibitem{Wallace2010Decoherence}
D. Wallace. `Decoherence and Ontology: or, How I Learned to Stop Worrying and Love FAPP'. In S. Saunders, J. Barrett, A. Kent and D. Wallace (eds.), \emph{Many Worlds? Everett, Quantum Theory, and Reality} (Oxford University Press, 2010)

\bibitem{Wallace2010}
D. Wallace. `How to Prove the Born Rule'. In S. Saunders, J. Barrett, A. Kent and D. Wallace (eds.), \emph{Many Worlds? Everett, Quantum Theory, and Reality} (Oxford University Press, 2010)

\bibitem{Wallace2011} 
D. Wallace. `The Everett Interpretation.' In R. Batterman (ed.) \emph{The Oxford Handbook of Philosophy of Physics} (Oxford University Press, 2013) 

\bibitem{Wallace2012}
D. Wallace. \emph{The Emergent Multiverse}. (Oxford University Press, 2012)

\bibitem{Wilson} 
A. Wilson. `Objective Probability in Everettian Quantum Mechanics.' \emph{British Journal of Philosophy of Science} (October 2012) 

\end{thebibliography}
\end{document}